\documentclass[prd,twocolumn]{revtex4}
\usepackage{amsmath}
\usepackage{graphicx,epsfig}
\usepackage{color}
\usepackage{mathrsfs}
\usepackage{setspace}
\usepackage{fancyhdr}

\newcommand{\be}{\begin{equation}}
\newcommand{\ee}{\end{equation}}
\newcommand{\bd}{\begin{equation*}}
\newcommand{\ed}{\end{equation*}}
\newcommand{\bea}{\begin{eqnarray}}
\newcommand{\eea}{\end{eqnarray}}
\newcommand{\bead}{\begin{eqnarray*}}
\newcommand{\eead}{\end{eqnarray*}}

\newcommand{\gapp}{\mathrel{\raise.3ex\hbox{$>$}\mkern-14mu
              \lower0.6ex\hbox{$\sim$}}}
\newcommand{\lapp}{\mathrel{\raise.3ex\hbox{$<$}\mkern-14mu
              \lower0.6ex\hbox{$\sim$}}}

\begin{document}
\pdfoutput=1

\title{Time Evolution of Temperature and Entropy of Various Collapsing Domain Walls}
\author{Evan Halstead}
\affiliation{HEPCOS, Department of Physics, SUNY at Buffalo, Buffalo, NY 14260-1500} 

\begin{abstract}
We investigate the time evolution of the temperature and entropy of gravitationally collapsing domain walls as seen by an asymptotic observer.  In particular, we seek to understand how topology and the addition of a cosmological constant affect the gravitational collapse.  Previous work has shown that the entropy of a spherically symmetric collapsing domain approaches a constant.  In this paper, we reproduce these results, using both a fully quantum and a semi-classical approach, then we repeat the process for a de Sitter Schwarzschild domain wall (spherical with cosmological constant) and a (3+1) BTZ domain wall (cylindrical).  We do this by coupling a scalar field to the background of the domain wall and analyzing the spectrum of radiation as a function of time.  We find that the spectrum is quasi-thermal, with the degree of thermality increasing as the domain wall approaches the horizon.  The thermal distribution allows for the determination of the temperature as a function of time, and we find that the late time temperature is very close to the Hawking temperature and that it also exhibits the proper scaling with the mass.  From the temperature we find the entropy.  Since the collapsing domain wall is what forms a black hole, we can compare the results to those of the standard entropy-area relation.  We find that the entropy does in fact approach a constant that is close to the Hawking entropy.  However, both the de Sitter Schwarzschild domain wall and the (3+1) BTZ domain wall show periods of decreasing entropy, which suggests that spontaneous collapse may be prevented.    
\end{abstract}
\maketitle

\section{Introduction}
It is well known, based primarily on the work of Bekenstein, Gibbons, and Hawking, that the entropy of a black hole is proportional to its area and that even though supposedly nothing can escape from within a black hole, quantum fluctuations near the event horizon will produce a spectrum of radiation that is thermal \cite{Hawking:1974sw, Hartle:1976tp, Gibbons:1976ue, Bekenstein:1973ur}.  Since then, much work has been done to reproduce this result with theories of quantum gravity, but most of these theories do not analyze the time evolution of the system.  The conventional process to determine the entropy uses the Bogolyubov method to first determine the temperature and from there find the entropy.  This, however, only utilizes the initial and final states of the system, and therefore there is no knowledge of the time dependence.  Recently, Vachaspati and Stojkovic developed a quantum treatment to determine the quantum radiation given off during gravitational collapse in a time dependent manner, then Greenwood expanded on this to determine the time dependence of the entropy \cite{Vachaspati:2006ki, Vachaspati:2007hr, Greenwood:2008vu, Greenwood:2008zg, Wang:2009ay, Greenwood:2010mr}.  These papers used a spherically symmetric collapsing domain wall for their analysis, and we wish to augment their work by seeing if a different topology and/or the existence of a cosmological constant will yield significantly different results.  Specifically, we wish to determine the time dependence of the entropy of a gravitationally collapsing de Sitter Schwarzschild domain wall (representing a Schwarzschild domain wall in a universe with a cosmological constant) and a (3+1) BTZ domain wall (representing a cylindrically symmetric domain wall).  Previous work (i.e. Ref.\cite{Greenwood:2009gp}) determined the equations of motion of these collapsing domain walls, and we will use those to complete the analysis of their thermodynamic properties.  To do this, we will first determine the wavefunctional of a scalar field coupled to the background of the collapsing shell.  Then, using the t=0 wavefunctional as a basis, we will determine the occupation number as a function of frequency.  This in turn will allow for the determination of the thermodynamic quantity $\beta$ and therefore the temperature.  We then determine the late time temperature as a function of horizon radius.  Using the thermodynamic definition of entropy, $dS=dQ/T$, we integrate to find entropy as a function of temperature, and since we know the temperature as a function of time we can find the entropy as a function of time.  We compare our results against the standard Hawking results, and we comment on the results.

\section{Fully Quantum Approach}
In this section we will outline the formalism (see \cite{Vachaspati:2007hr}) for determining the time evolution of the temperature and entropy of the collapsing domain wall.  We will keep the analysis in terms of a general metric, then the following sections will use the specific metrics to see how the details differ.  First, we will consider the radiation given off by the domain wall during gravitational collapse.  This will be done by coupling a scalar field to the background of the collapsing shell.  We will use the t=0 wavefunction of the scalar field as a basis, which will allow us to determine the occupation number for each frequency at later times.  For all three metrics, the plot of occupation number versus frequency at different time slices mimics that of a Planck distribution.  This allows for the determination of the temperature as a function of time.  Then we use the thermodynamic definition of entropy as it relates to temperature to determine the time evolution of the entropy.

We will decompose the scalar field as
\begin{equation}
\Phi = \sum_{k} a_k(t)u_k(r).
\end{equation}
The exact form of $u_k(r)$ will not be important to us.  For a general metric, we will take the exterior metric as
\begin{align}
  (ds^2)_+=&-f(r)dt^2+\frac{1}{f(r)}dr^2+r^2d\Omega^2 \hspace{2mm} \text{for $r>R(t)$}
       \label{out_metric}
\end{align}
and the interior metric as
\be
  (ds^2)_-=-A(r)dT^2+dr^2+\frac{1}{A(r)}r^2d\Omega^2 \hspace{2mm} \text{for $r<R(t)$}
  \label{in_metric}
\ee
where 
\be
d\Omega^2=d\theta^2+\sin{\theta}^2d\phi^2
\ee
for the Schwarzschild and de Sitter Schwarzschild metrics and
\be
d\Omega^2=d\phi^2+dz^2
\ee
for the (3+1) BTZ metric.
To find the modes $a_k(t)$, we will insert the metrics given by Eqs.(\ref{out_metric}) and (\ref{in_metric}) into the action 
\begin{equation}
S_\Phi = \int d^4x \sqrt{-g} g^{\mu\nu} \partial_\mu \Phi \partial_\nu \Phi,
\end{equation}  
where we are using $r=R(t)$ as the domain wall's position.
The total action will be written as the sum
\begin{equation}
S = S_{in} + S_{out},
\end{equation}
where 
\begin{equation}
S_{in} = 2\pi \int dt \int_0^{R(t)}dr r^2 \left (-\frac{(\partial_t \Phi)^2}{\dot{T}} + \dot{T}(\partial_r \Phi)^2\right )
\end{equation}
is the metric inside the shell, 
\begin{equation}
S_{out} = 2\pi \int dt \int_{R(t)}^\infty dr r^2 \left (-\frac{(\partial_t\Phi)^2}{f}+f(\partial_r\Phi)^2\right)
\end{equation}
is the metric outside the shell, and
\begin{equation}
\frac{dT}{dt} = \frac{1}{A} \sqrt {Af-\frac{\left [A-f\right] \dot{R}^2}{f}}.
\end{equation}
According to Ref.\cite{Greenwood:2009gp}, 
\begin{equation}\label{eq:Rdotnoapprox}
\dot{R} = f \sqrt{1-\frac{fR^4}{h^2}}
\end{equation}
where 
\begin{equation}
h = \frac{f^{3/2}R^2}{\sqrt{f^2-\dot{R}^2}},
\end{equation}
so $dT/dt$ can be rewritten as 
\begin{equation}
\frac{dT}{dt} = \frac{f}{A}\sqrt{1+\left [ A - f \right ] \frac{R^4}{h}}.
\end{equation}
In the region of interest, where $R\rightarrow R_H$ and therefore $f\rightarrow 0$, the kinetic term in $S_{in}$ dominates over that in $S_{out}$ while the gradient term in $S_{in}$ is subdominant to that in $S_{out}$.  This yields the approximate action
\begin{equation}
S \approx 2\pi \int dt \left [ -\int_0^{R_H} dr r^2 \frac{(\partial_t \Phi)^2}{f} + \int_{R_H}^{\infty} dr r^2 f (\partial_r \Phi)^2 \right ].  
\end{equation}
Substituting the expansion for the scalar field $\Phi$ produces 
\begin{eqnarray} \label{eq:scalaraction}
&S& = \int dt \left[ -2\pi \int_0^{R_H} dr r^2 \frac{\sum_{k,k'}^{} \dot a_k(t)\dot a_{k'}(t)u_k(r)u_{k'}(r)}{f} \right . \nonumber \\
&+&  \left . 2\pi \int_{R_H}^\infty dr r^2 f \sum_{k,k'}a_k(t)a_{k'}(t)u'_k(r)u'_{k'}(r)\right ].
\end{eqnarray}
  Now let
\begin{equation}
M_{kk'} = 4\pi \int_0^{R_H} dr r^2 u_k(r) u_{k'}(r)
\end{equation}  
  and
\begin{equation}
N_{kk'} = - 4\pi \int_{R_H}^\infty dr r^2 f(r) u'_k(r) u'_{k'}(r),
\end{equation}   
 which allows us to rewrite $(\ref{eq:scalaraction})$ as
 \begin{equation} \label{eq:scalaraction1}
 S = \sum_{k,k'} \int dt \left [ -\frac{1}{2f}\dot a_k M_{kk'} \dot a_{k'} - \frac{1}{2} a_k N_{kk'} a_{k'}\right ].
 \end{equation}
   The action $(\ref{eq:scalaraction1})$ corresponds to the Hamiltonian
   \begin{equation}
   H = \frac{1}{2}f\Pi_k M_{kk'}^{-1} \Pi_{k'} + \frac{1}{2} a_k N_{kk'} a_{k'},
   \end{equation}
 where
 \begin{equation}
 \Pi = \frac{\partial L}{\partial \dot a}.
 \end{equation}
 Since $\textbf M$ and $\textbf N$ are Hermitian matrices, it is possible to do a principle axis transformation to diagonalize them simultaneously (see for example section 6.2 of \cite{Goldstein}).  The exact form of them, however, is not important here.  The Schr$\mathrm{\ddot {o}}$dinger Equation for a single mode will then be
\begin{equation} \label{eq:hamiltonian}
\left [-\frac{1}{2m}f\frac{\partial^2}{\partial b^2}+\frac{1}{2} K b^2 \right ] \psi (b,t) = i \frac{\partial \psi (b,t)}{\partial t}, 
\end{equation}
where $m$ and $K$ are the eigenvalues of $\textbf M$ and $\textbf N$, respectively, and $b$ is the eigenmode.

\begin{figure}
\centering
\includegraphics{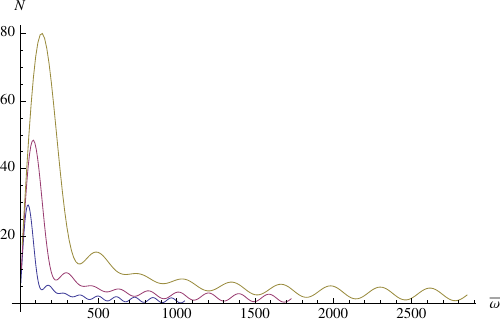}
\caption{Occupation number as a function of frequency for a Schwarzschild domain wall quantized in $R$.  From bottom to top, the three curves represent time slices for $t/R_s$=13, 14, and 15, respectively.}
\label{fig:Nschwquantum}
\end{figure}

\begin{figure}
\centering
\includegraphics{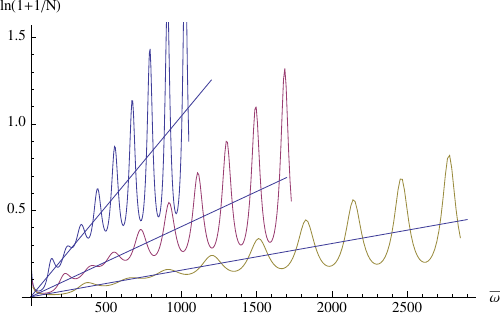}
\caption{The slope of this plot represents the thermodynamic quantity $\beta$ for a Schwarzschild domain wall quantized in $R$.  From top to bottom, the three curves represent time slices for $t/R_s$=13, 14, and 15, respectively.  The best fit lines are also included.}
\label{fig:lnNschwquantum}
\end{figure}

\begin{figure}
\centering
\includegraphics{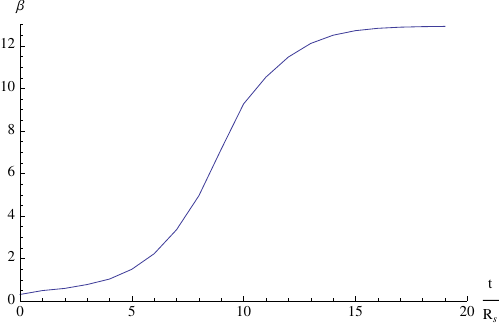}
\caption{$\beta$ as a function of time for a Schwarzschild domain wall quantized in $R$.}
\label{fig:betaschwquantum}
\end{figure}

\begin{figure}
\centering
\includegraphics{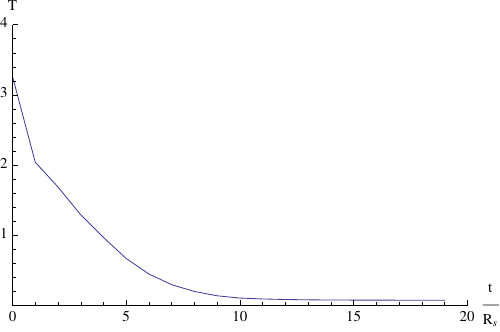}
\caption{Temperature as a function of time for a Schwarzschild domain wall quantized in $R$.}
\label{fig:Tschwquantum}
\end{figure}

It should be noted at this point that this represents only the Hamiltonian associated with the radiation.  The total Hamiltonian of the system is the Hamiltonian of the radiation plus that of the domain wall.
\be
H_{total}=H_{rad}+H_{wall}
\ee
The Hamiltonian of the wall is given by (see Ref.\cite{Greenwood:2009gp})
\be
  H\approx\frac{2\pi\mu f^{3/2}R^2}{\sqrt{f^2-\dot{R}^2}}=\sqrt{(f\Pi)^2+f(2\pi\mu R^2)^2}.
  \label{AS_apH}
\ee
 We will specifically work in the limit $R\rightarrow R_H$, which corresponds to $f\rightarrow 0$.  In this limit, the second term is subdominant since $\Pi\sim f^{-3/2}$ and
\be
H_{wall}\approx -f\Pi,
\ee
where we have chosen the negative sign because we are looking for the collapsing solution.  Therefore, taking the wall into account, the Schr$\mathrm{\ddot {o}}$dinger Equation takes the form
\be
\label{eq:hamiltoniantotal}
if\frac{\partial \Psi}{\partial R}-\frac{f}{2m}\frac{\partial^2 \Psi}{\partial b^2}+\frac{K}{2}b^2\Psi=i\frac{\partial \Psi}{\partial t}.
\ee
As an ansatz, we will look for stationary solutions.
\be
\Psi(b,R,t)=e^{-iEt}\psi(b,R)
\label{Psipsi}
\ee
With this ansatz, the time-independent Schr$\mathrm{\ddot {o}}$dinger Equation becomes
\be
-\frac{1}{2m}\frac{\partial^2\psi}{\partial b^2}+\frac{m}{2}\omega^2 b^2\psi-\epsilon \psi=-i\frac{\partial \psi}{\partial R},
\ee
where
\be
\omega^2=\frac{K}{mf}=\frac{\omega_0^2}{f}
\ee
and
\be
\epsilon=\frac{E}{f}.
\ee
We will now assume a solution of the form
\begin{eqnarray}
\psi(b,R) &=& Exp[-i\int^R\epsilon(R')dR']\phi(b,R)  \nonumber \\
&=&  Exp[-iEu]\phi(b,R),
\label{psiphi}
\end{eqnarray}
where 
\be
u=\int_{R_0}^R \frac{dR'}{f(R')}.
\label{utoR}
\ee
The Schr$\mathrm{\ddot {o}}$dinger Equation now becomes
\be
-\frac{1}{2m}\frac{\partial^2 \phi}{\partial b^2}+\frac{m\omega^2}{2}b^2\phi=i\frac{\partial\phi}{\partial\eta},
\ee
where
\be 
\eta=R_H-R.
\ee
The solution is 
\be
\phi(b,\eta)=e^{i\alpha(\eta)}\left(\frac{m}{\pi\rho^2}\right)^{1/4}Exp\left[\frac{im}{2}\left(\frac{\rho_{\eta}}{\rho}+\frac{i}{\rho^2}\right)b^2\right],
\ee
where $\rho_{\eta}$ is the derivative of $\rho(\eta)$ with respect to $\eta$.  The function $\rho(\eta)$ comes from solving the differential equation
\be
\rho_{\eta\eta}+\omega^2(\eta)\rho=\frac{1}{\rho^3}
\ee
where
\be
\omega^2(\eta)=-\frac{KR}{m\eta}\approx-\frac{KR_H}{m\eta}.
\ee
Here we are again assuming that we are in the limit $R\rightarrow R_H$.
To solve the differential equation for $\rho(\eta)$, we will use the initial conditions
\be
\rho(\eta_i)=\frac{1}{\omega(\eta_i)}, \rho_{\eta}(\eta_i)=0.
\ee
Finally, the phase $\alpha$ is defined by
\be
\alpha(\eta)=-\frac{1}{2}\int^{\eta}\frac{d\eta'}{\rho^2(\eta')}.
\ee
Combining Eqs. (\ref{Psipsi}) and (\ref{psiphi}) produces
\be
\Psi(b,R,t)=e^{-iE(u+t)}\phi(b,\eta),
\ee
which we can then use to construct wavepackets by superposing stationary solutions given by
\be
\Psi(b,R,t)=\frac{1}{(\pi \sigma^2)^{1/4}}e^{-(u+t)^2/2 \sigma^2}\phi(b,\eta),
\ee
where $\sigma$ is the width of the wavepacket and where the prefactor is due to normalization in the $u$ coordinate.  This solution describes the wavefunction of the domain wall as it moves toward the horizon $R_H$ ($u=-\infty$).  The function $\phi(b,\eta)$ describes the radiation from the domain wall, and together these will enable us to find the temperature and entropy.
To proceed from here, we wish to decompose the wavefunction $\Psi$ into a set of basis wavefunctions which we will denote as $\phi_n$.  In the semiclassical case, which we will describe later, the occupation number is given by
\be
N=\sum_n n\big|\langle \phi_n \big|\Psi\rangle\big|^2.
\ee
Presently, we are utilizing a complete quantum treatment where the wavefunction is quantized in $R$, however.  This means that the occupation number $N$ is a function of $R$, so to obtain the occupation number only as a function of time we need to find the expectation value by integrating the probability distribution over position, which in this case is represented by the scaled position coordinate $u$.  
\be
\langle N(t)\rangle=\int du \langle N(R,t)\rangle=\int du \sum_n n\big|\langle\phi_n\big|\Psi\rangle\big|^2
\label{Nexp}
\ee
To proceed with the integration, we will choose our basis wavefunctions to be simple harmonic oscillator states given by
\be
\phi_n(b)=\left(\frac{m\bar{\omega}}{\pi}\right)^{1/4}\frac{e^{-m\bar{\omega}b^2/2}}{\sqrt{2^nn!}}H_n(\sqrt{m\bar{\omega}}b),
\ee
where $H_n$ are Hermite polynomials and $\bar{\omega}$ is the expected value of the frequency given by
\begin{eqnarray}
\bar{\omega}&=&\langle \omega \rangle=\int du \frac{1}{\sqrt{\pi\sigma^2}}e^{-(u+t)^2/\sigma^2}\omega \nonumber \\ &=&\int du \frac{1}{\sqrt{\pi\sigma^2}}e^{-(u+t)^2/\sigma^2} \sqrt{\frac{K}{mf}}.
\end{eqnarray}
Then
\begin{eqnarray}
\langle\phi_n\big|\Psi\rangle &=& \frac{1}{(\pi \sigma^2)^{1/4}}e^{-(u+t)^2/2\sigma^2}\frac{(-1)^{n/2}e^{-i\alpha}}{(\bar{\omega}\rho^2)^{1/4}} \nonumber \\ &\times &\sqrt{\frac{2}{P}} \left(1-\frac{2}{P}\right)^{n/2}\frac{(n-1)!!}{\sqrt{n!}},
\end{eqnarray}
where
\be
P=1-\frac{i}{\bar{\omega}}\left(\frac{\rho_{\eta}}{\rho}+\frac{i}{\rho^2}\right).
\ee
The sum in Eq. (\ref{Nexp}) can be done explicitly.
\begin{eqnarray}
\langle N(t)\rangle=&\int & du \frac{1}{\sqrt{\pi\sigma^2}}e^{-(u+t)^2/\sigma^2}\frac{\bar{\omega}\rho^2}{4} \nonumber \\ &\times& \left[\left(1-\frac{1}{\bar{\omega}\rho^2}\right)^2+\left(\frac{\rho_{\eta}}{\bar{\omega}\rho}\right)^2\right]
\label{Nexpfinal}
\end{eqnarray}
We now wish to determine the occupation number as a function of frequency $\bar{\omega}$. To determine the occupation number of a given frequency $\bar{\omega}$ at a given time, we must first pick a value for $\omega_0$ and solve the differential equation for $\rho(\eta)$.  Then we insert that solution into Eq. (\ref{Nexpfinal}) and integrate over $u$.  Then the plot of occupation number versus frequency is created by iterating this process over a wide range of values of $\omega_0$.  We will find that in each case, the plot is very similar to a typical Planck distribution.  Therefore, we can find the temperature of the radiation by treating the occupation number of each eigenmode $b$ as if it followed the Plank distribution,
\begin{equation}
N_P=\frac{1}{e^{\beta \bar{\omega}}-1}.
\end{equation}
If one plots $\ln{(1+1/N)}$ as a function of $\bar{\omega}$, the slope will yield $\beta$.  Simply inverting $\beta$, however, will not yield the temperature in the asymptotic observer's time since the frequency we found is the frequency in the scaled coordinate $u$.  Therefore, the temperature according to the asymptotic observer can be found by rescaling back.  This is given by
\be
\beta^{(t)}=\frac{\beta}{f}.
\label{betascale}
\ee
Since $f$ is a function of $R$, and the position of the domain wall $R$ is being described by a quantum wavefunction, then we will determine $f$ by using the expectation value of $R$,
\be
\langle R \rangle=\int du \frac{1}{\sqrt{\pi\sigma^2}}e^{-(u+t)^2/\sigma^2} R  
\ee
where the value of $R$ corresponding to each $u$ is determined from Eq.(\ref{utoR}).
For a spherically symmetric domain wall, we use the Schwarzschild metric, which gives 
\be
u=R+R_sln\big|\frac{R}{R_s}-1\big|-R_0-R_sln\big|\frac{R}{R_s}-1\big|.
\ee
The corresponding plot of occupation number versus expected frequency is shown in Fig.(\ref{fig:Nschwquantum}).
Notice, as stated earlier, that the plot resembles that of a Planck distribution.  In fact, the distribution gets closer to a Planck distribution as time goes on as can be seen from the fact that the low frequency occupation number gets larger and larger over time.  The plot of $ln(1+1/N)$ is shown in Fig.(\ref{fig:lnNschwquantum}).
A perfect Planck distribution would show up as a straight line, so we have superimposed a best fit line on each curve.  It is clear from this plot that the domain wall has not yet thermalized; therefore we will call this a quasi-thermal state.  Though the deviations from thermality are large at points, we wish to point out two things: 1) the largest deviations from a thermal distribution occur at larger frequencies, which are the last frequencies to thermalize due to their relatively small number, and 2) as time goes on, the amplitude of the oscillations dampens out, and therefore the late time behavior approaches that of a perfect blackbody.  By finding the slope of these lines at regular intervals of time and rescaling to the asymptotic observer's coordinates through Eq.(\ref{betascale}), we produce a plot of $\beta^{(t)}$ as a function of time, shown in Fig.(\ref{fig:betaschwquantum}).
Finally, we invert $\beta^{(t)}$ to find the temperature as a function of time, as shown in Fig.(\ref{fig:Tschwquantum}).
Notice from Fig.(\ref{fig:Tschwquantum}) that as the wall approaches the horizon radius, the temperature approaches a constant.  We will now compare this temperature to the well known Hawking temperature given by
 \begin{equation}
 T_H=\frac{\kappa}{2\pi}
 \end{equation}
 where $\kappa$ is the surface gravity, which for a Schwarzchild black hole is 
 \begin{equation}
 \kappa=\frac{1}{2R_s}.
 \end{equation}
 For $R_s=1$, the ratio of the late time temperature to the accepted Hawking temperature is $T/T_H=0.973$. 
 
 \section{Semi-classical Approach}
We will now repeat the previous process using a semi-classical approach.  The reason for this is because, as we will see, the results are qualitatively very similar to the fully quantum treatment while the mathematical analysis is significantly less involved.  The biggest difference between the two methods is that the amplitude of the oscillations in the semi-classical case are much smaller, which will aid in the determination of the temperature.  We start by revisiting the Hamiltonian of the total system, given by Eq.(\ref{eq:hamiltoniantotal}).  This time we will insert the classical equation of motion for $R(t)$, which in the limit $R\rightarrow R_H$ is approximately given by (see \cite{Greenwood:2008vu})
 \begin{equation}\label{eq:Rdot}
\dot R \approx -f,
\end{equation}
where again the negative sign is chosen because we are choosing to examine the collapsing solution.  Since $R(t)$ is only a function of time, we can rewrite Eq.($\ref{eq:hamiltoniantotal}$) as 
\begin{equation}
if\frac{1}{\dot R}\frac{\partial \Psi}{\partial t}-\frac{f}{2m}\frac{\partial^2\Psi}{\partial b^2}+\frac{K}{2}b^2\Psi=i\frac{\partial \Psi}{\partial t}.
\end{equation} 
According to Eq.(\ref{eq:Rdot}), however, $f/\dot{R}=-1$, therefore
\begin{equation}
-\frac{f}{2m}\frac{\partial^2\Psi}{\partial b^2}+\frac{K}{2}b^2\Psi = 2i\frac{\partial \Psi}{\partial t}.
\label{eq:quantumhamiltonian}
\end{equation}
   
We rewrite Eq.(\ref{eq:quantumhamiltonian}) in the standard simple harmonic oscillator form
\begin{equation}
\left [ -\frac{1}{2m}\frac{\partial^2}{\partial b^2}+\frac{m}{2}\omega^2 b^2\right ] \psi (b,\tilde{\eta}) = 2i \frac{\partial \psi (b,\tilde{\eta})}{\partial \tilde{\eta}}, 
\end{equation}
where 
\begin{eqnarray}
\tilde{\eta} = \frac{1}{2}\int_0^t dt' f \\
\omega^2 = \frac{K}{fm}=\frac{\omega_0^2}{f},
\label{eq:ttoeta}
\end{eqnarray}
and where we have chosen $\tilde{\eta}(t=0)=0$.

At early times, the initial vacuum state for the wavefunction $\psi$ is that of a simple harmonic oscillator given by
\begin{equation}
\psi (b, \tilde{\eta} =0) = \left (\frac{m \omega_0}{\pi}\right )^{1/4} e^{-m\omega_0 b^2/2}.  
\end{equation}
At later times, the exact solution for the wavefunction is (see Ref.\cite{Dantas:1990rk})
\begin{equation} \label{eq:latewavefunction}
\psi (b, \tilde{\eta}) = e^{i\alpha (\tilde{\eta})}\left (\frac{m}{\pi \rho^2}\right)^{1/4} \exp{\left [i\frac{m}{2}\left(\frac{\rho_{\tilde{\eta}}}{\rho}+\frac{i}{\rho^2}\right)b^2\right ]},
 \end{equation}
where $\rho_{\tilde{\eta}}$ denotes the derivative of $\rho(\tilde{\eta})$ with respect to $\tilde{\eta}$, and $\rho(\tilde{\eta})$ is the solution to the equation
\begin{equation}
\frac{\partial^2 \rho}{\partial \tilde{\eta}^2}+\omega^2(\tilde{\eta})=\frac{1}{\rho^3}
\end{equation}
 with initial conditions
 \begin{eqnarray}
 \rho(0)=\frac{1}{\sqrt{\omega_0}} \\
 \rho_{\tilde{\eta}}(0)=\frac{\partial \rho}{\partial \tilde{\eta}} \mid_0 = 0.
 \end{eqnarray}
The phase $\alpha$ is given by
\begin{equation}
\alpha(\tilde{\eta})=-\frac{1}{2}\int_0^{\tilde{\eta}} \frac{d\tilde{\eta'}}{\rho^2(\tilde{\eta'})}.
\end{equation}   

\section{Occupation Number of the Radiation}

Consider an observer with detectors that are designed to register particles for the scalar field $\phi$ at early times.  At late times, the observer will interpret each mode $b$ in terms of the simple harmonic oscillator states, with final frequency $\bar{\omega}$.  The number of quanta in eigenmode $b$ can be found by decomposing the wavefunction $(\ref{eq:latewavefunction})$ into the simple harmonic oscillator states and evaluating the occupation number.  The wavefunction $\psi$ written in terms of the simple harmonic basis, $\{\varphi_n \}$, at $t=0$ is given by
\begin{equation}
\psi (b,t) = \sum_n c_n(t) \varphi_n(b),
\end{equation}
where 
\begin{equation}
c_n(t) = \int db \varphi_n^\ast(b) \psi (b,t),
\end{equation} 
which is the overlap of a Gaussian with the simple harmonic basis functions.  The occupation number at eigenfrequency $\bar \omega$ is given by the expectation value
\begin{equation}
N(t,\bar \omega) = \sum_n n |c_n|^2.  
\end{equation}
After substitution, we find that the occupation number in the eigenmode $\textit b$ is given by (see Appendix B in Ref.\cite{Greenwood:2009pd})
\begin{equation}
N(t,\bar{\omega}) = \frac{\bar{\omega} \rho^2}{4}\left [ \left (1-\frac{1}{\bar{\omega} \rho^2}\right )^2 + \left (\frac{2\rho_t}{f\bar{\omega} \rho}\right )^2\right ].
\end{equation}
 We will find that in each case, just as with the quantum treatment, the plot is very similar to a typical Planck distribution.  Therefore, we can find the temperature of the radiation by treating the occupation number of each eigenmode $b$ as if it followed the Plank distribution,
\begin{equation}
N_P=\frac{1}{e^{\beta \bar{\omega}}-1}.
\end{equation}
If one plots $\ln{(1+1/N)}$ as a function of $\bar{\omega}$, the slope will yield $\beta$.
It should be noted at this point that $\beta$ is in terms of scaled time $\tilde{\eta}$ instead of the asymptotic observer time $t$.  Therefore, we must rescale $\beta$ back to the observer's time using Eq.(\ref{eq:ttoeta}), which is achieved by
\begin{equation}
\beta^{(t)}=\frac{\beta^{(\tilde{\eta})}}{f}.
\end{equation}
Finally, the temperature as a function of the observer's time can be found by inverting $\beta$.
\begin{equation}
T(t)=\frac{1}{\beta^{(t)}}
\end{equation}

 \section{Entropy}
 The thermodynamic definition of entropy in terms of temperature is
 \begin{equation}
 S=\int{\frac{dQ}{T}}.
 \end{equation}
 Since changing the energy of the domain wall is the same as changing the mass, this can also be written as 
 \begin{equation}
 S=\int{\frac{dM}{T}}.
 \end{equation}
It should be noted that this represents the entropy only of the domain wall, not of the total system consisting of the domain wall plus the radiation.  This is due to the fact that the mass we will be using in the entropy equation is actually the mass as it depends on the radius of the domain wall (i.e. $M=R_S/(2G)$ for Schwarzschild).  The mass of the entire system, however, is a conserved quantity and is therefore independent of temperature.  Using this information, if we can determine how the temperature of the domain wall depends on the mass, we will be able to find an expression for the entropy in terms of the temperature.  And since we know the time evolution of the temperature, we can determine the time evolution of the entropy as well.  
  
 \subsection{Schwarzschild Domain Wall}
\begin{figure}
 \centering
 \includegraphics[width=3.2in]{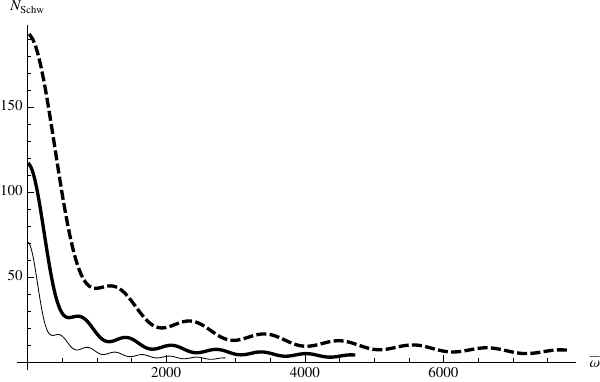}
 \caption{Occupation number as a function of frequency for a semi-classical Schwarzschild domain wall.  From bottom to top, the three curves represent time slices for $t/R_s$=13 (solid), 14 (bold), and 15 (dashed), respectively.} 
 \label{fig:Ntotalschw}
 \end{figure}

\begin{figure}
 \centering
 \includegraphics[width=3.2in]{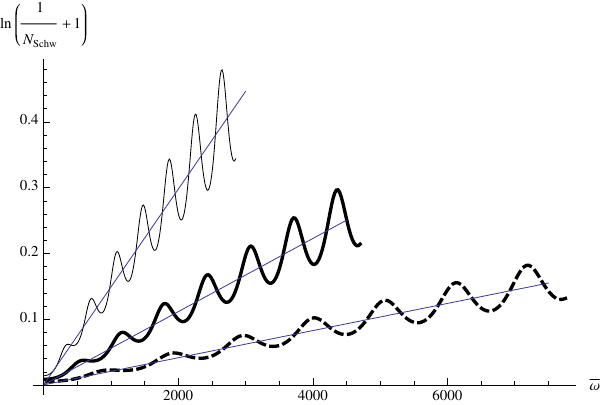}
 \caption{The slope of this plot represents the thermodynamic quantity $\beta$ for a semi-classical Schwarzschild domain wall.  From top to bottom, the three curves represent time slices for $t/R_s$=13 (solid), 14 (bold), and 15 (dashed), respectively.  The best-fit curves are also included.}
 \label{fig:lnNtotalschw}
 \end{figure}

\begin{figure}
 \centering
 \includegraphics{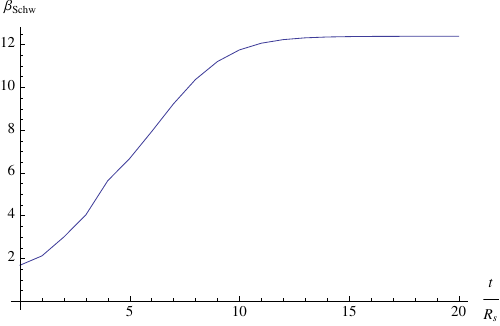}
 \caption{$\beta$ as a function of time for a semi-classical Schwarzschild domain wall.}
 \label{fig:betatotalschw}
 \end{figure}

 For the Schwarzchild metric,
 \begin{equation}
 f_{Schw}=1-\frac{R_s}{R}.
 \end{equation}
 According to Ref.\cite{Greenwood:2008vu}, the classical equation of motion of the domain wall from the point of view of an asymptotic observer is
 \begin{equation}
 R_{Schw}(t)=R_s+(R_0-R_s)e^{-t/R_s}.
 \end{equation}
 For the total system, the plot of occupation number as a function of frequency is shown in Fig.(\ref{fig:Ntotalschw}).  Notice, as stated earlier, how similar this is to a Planck distribution.  In fact, as time goes on, the distribution becomes more and more Planck-like as the oscillations in the curve dampen out and the $\bar{\omega}=0$ occupation number gets larger.  The plot of $ln(1+1/N_{Schw})$ as a function of frequency is shown in Fig.(\ref{fig:lnNtotalschw}).  The straight lines are best-fit lines whose slopes represent $\beta$ at the chosen time.  While in scaled time $\tilde{\eta}$ the $\beta$ actually diverges, in the asymptotic observer's time $t$ the $\beta$ actually approaches a constant, as shown in Fig.(\ref{fig:betatotalschw}).
 
\begin{figure}
 \centering
 \includegraphics{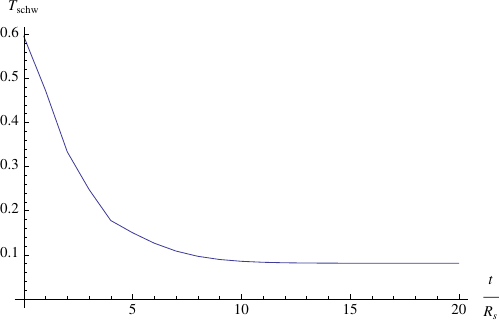}
 \caption{Temperature as a function of time for a semi-classical Schwarzschild domain wall.}
 \label{fig:temptotalschw}
 \end{figure}

 The corresponding plot of the temperature as a function of time is shown in Fig.(\ref{fig:temptotalschw}).
 Notice that as the wall approaches the horizon radius, the temperature approaches a constant.  We will now compare this temperature to the well known Hawking temperature as was done previously in the quantum treatment.  For $R_s=1$, the ratio of the temperature to the Hawking temperature $T/T_H=0.998$.  
 
\begin{figure}
 \centering
 \includegraphics{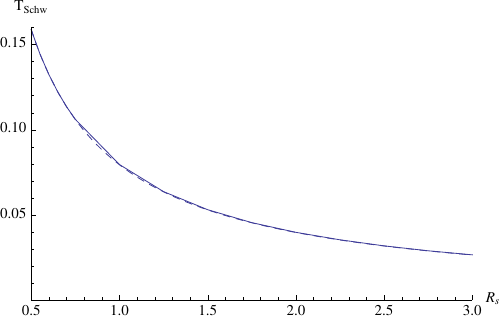}
 \caption{Temperature as a function of $R_s$ for a semi-classical Schwarzschild domain wall.  The dashed line represents the Hawking model.}
 \label{fig:tempdepschw}
 \end{figure}

\begin{figure}
 \centering
 \includegraphics{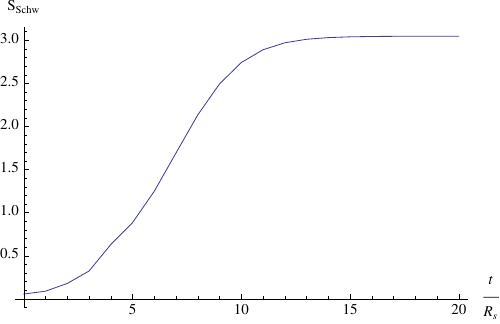}
 \caption{Entropy as a function of time for a semi-classical Schwarzschild domain wall.}
 \label{fig:entropyschw}
 \end{figure}

 We will now show that the temperature of the domain wall also exhibits the proper scaling with the Schwarzschild radius by plotting the late time temperature as a function of $R_s$.  Fig.(\ref{fig:tempdepschw}) shows this plot for $R_s$ ranging from 0.5 to 3.  
 A best-fit curve proportional to $1/R_s$ has been overlayed on the plot to show that it exhibits the expected scaling with temperature.  The equation of the best-fit curve is 
 \begin{equation}
 T_{Schw}=\frac{0.0794262}{R_s}.
 \end{equation}  
 What this plot allows us to do now is to determine how the entropy depends on the temperature.  As stated earlier, the thermodynamic definition of entropy in terms of temperature is
 \begin{equation}
 S=\int \frac{dQ}{T}
 \end{equation}
 Since changing heat is equivalent to changing mass, $dQ=dM=dR_s/(2G)$.  We will write the temperature as 
 \begin{equation}
 T_{Schw}=\frac{\gamma_{Schw}}{R_s},
 \end{equation}
 where $\gamma_{Schw}=0.0794262$.
 After integrating and solving for entropy in terms of temperature, we find
 \begin{equation}
 S_{Schw}=\frac{\gamma}{4T_{Schw}^2}.
 \end{equation}
  Since we have a plot of temperature versus time, this allows us to find the entropy versus time, as shown in Fig.(\ref{fig:entropyschw}).
As expected, the entropy approaches a constant as the domain wall approaches the horizon.
 Once again, we can compare the late-time entropy to the well-known result
 \begin{equation}
 S_H=\frac{A}{4}.
 \end{equation}
In the case of a Schwarzschild black hole with $R_s=1$, the ratio of the entropy to the Hawking entropy is $S_{Schw}/S_H=0.970$.  The fact that this ratio is not equal to one is not necessarily an indication of an inconsistency with the Hawking result, but rather is most likely the result of the numerous approximations that were required to establish this formalism as well as the arbitrary choices of "late times".  If anything, this result should be interpreted as yet another confirmation of the Hawking results.  

\subsection{de Sitter Schwarzschild Domain Wall}
\begin{figure}
\centering
\includegraphics{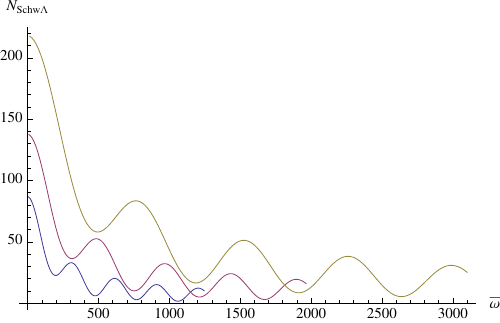}
\caption{Occupation number as a function of frequency for a de Sitter Schwarzschild domain wall.  From bottom to top, the three curves represent time slices for $t=13$ (solid), $t=14$ (bold), and $t=15$ (dashed).}  
\label{fig:Nschwcosmo}
\end{figure}

\begin{figure}
\centering
\includegraphics{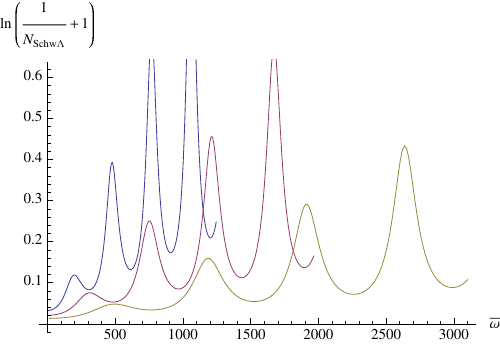}
\caption{The slope of this plot represents the thermodynamic quantity $\beta$ for a de Sitter Schwarzschild domain wall.  From top to bottom, the three curves represent time slices for $t=13$ (solid), $t=14$ (bold), and $t=15$ (dashed).}
\label{fig:lnNschwcosmo}
\end{figure}

\begin{figure}
\centering
\includegraphics{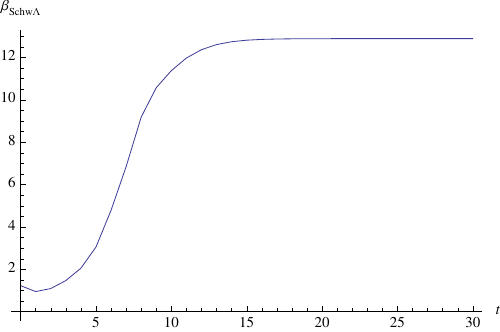}
\caption{$\beta$ as a function of time for a de Sitter Schwarzschild domain wall.}
\label{fig:betaschwcosmo}
\end{figure}

\begin{figure}
\centering
\includegraphics{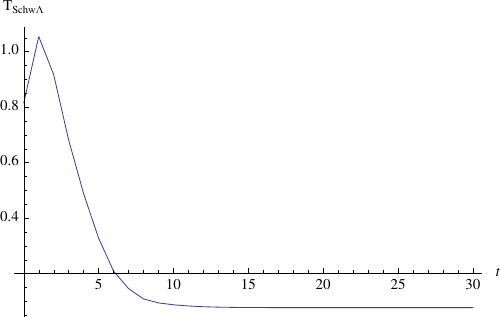}
\caption{Temperature as a function of time for a de Sitter Schwarzschild domain wall.}
\label{fig:tempschwcosmo}
\end{figure}

\begin{figure}
\centering
\includegraphics{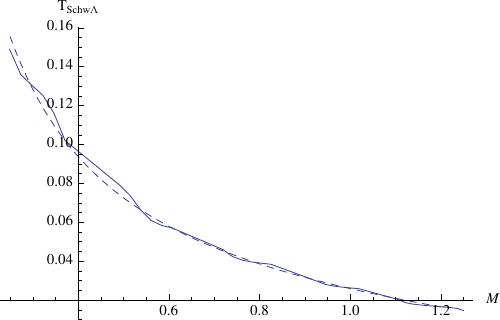}
\caption{Temperature as a function of mass for a de Sitter Schwarzschild domain wall.  The dashed line represents the Hawking model.}
\label{fig:tempdepschwcosmo}
\end{figure}

\begin{figure}
\centering
\includegraphics{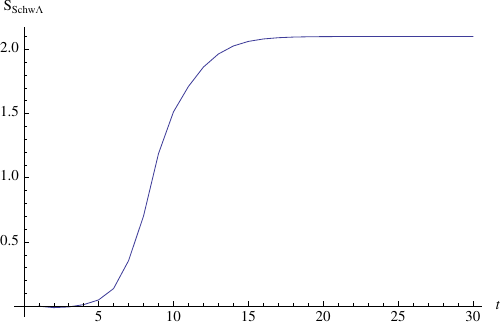}
\caption{Entropy as a function of time for a de Sitter Schwarzschild domain wall.}
\label{fig:entropyschwcosmo}
\end{figure}

We will now repeat this procedure for the de Sitter Schwarzschild domain wall, where
\be
f_{Schw\Lambda}=1-\frac{2GM}{R}-\frac{\Lambda R^2}{3}.
\ee
Notice that this particular metric has two horizons: one is the usual Schwarzschild horizon and the other is a cosmological horizon from the cosmological constant. The corresponding plots of occupation number, $ln(1+1/N)$, $\beta$, and temperature are shown in Fig.(\ref{fig:Nschwcosmo}), Fig.(\ref{fig:lnNschwcosmo}), Fig.(\ref{fig:betaschwcosmo}), and Fig.(\ref{fig:tempschwcosmo}), respectively.  The input values used for the plots are $R_H=1$ and $\Lambda=0.0557$.  This particular value of the cosmological constant was chosen because it put the initial position of the domain wall just inside the cosmological horizon.
Notice how, from Fig.(\ref{fig:tempschwcosmo}), there is a short time where the temperature increases before it starts to decrease again.  This is an effect which is much more prominent when the domain wall starts just inside the cosmological horizon, and it is suppressed the smaller the cosmological constant is.  
According to \cite{Arraut:2008hc}, the Hawking temperature of a de Sitter black hole is
\be
T_H=\frac{1}{8\pi M}-\frac{\Lambda M}{2\pi}.
\ee  
From this we find that the ratio of the late time temperature in Fig.(\ref{fig:tempschwcosmo}) to the accepted Hawking temperature is $T_{Schw\Lambda}/T_H=1.02$.  
As before, we will determine how the late time temperature scales with the mass of the domain wall.  Note that earlier we found $T$ as a function of $R_s$ because the relationship between $R_s$ and $M$ was trivial, $R_s=2GM$.  This time, we will find temperature as a function of $M$ because the relationship between the mass and the horizon radius is nontrivial, specifically
\be
M=\frac{R_H}{2}\left(1-\frac{\Lambda R_H^2}{3}\right).
\ee
The plot of the domain wall's late time temperature as a function of mass, along with the best-fit curve, is shown in Fig.(\ref{fig:tempdepschwcosmo}).
The equation of the best-fit curve is
\be
T_{Schw\Lambda}=\frac{0.0397304}{M}-0.0138729 M.
\ee
From here, we integrate to find the entropy,
\bea
S&=&\frac{\gamma_{Schw\Lambda}}{12 \nu^3(T_{Schw\Lambda}-\nu)} \left [ \frac{3\gamma_{Schw\Lambda}^2\nu\Lambda}{T_{Schw\Lambda}-\nu}-6\gamma_{Schw\Lambda}^2\Lambda \right.  \nonumber \\ 
 &&\left. + \nu^2\left(6-\frac{2\gamma_{Schw\Lambda}^2\Lambda}{(T_{Schw\Lambda}-\nu)^2}\right ) \right ]  \nonumber \\
  &+&  \frac{1}{2\nu^4}\left(-\gamma_{Schw\Lambda}\nu^2+\gamma_{Schw\Lambda}^3\Lambda\right) \nonumber \\
  & \times& ln\left(\gamma_{Schw\Lambda}+\frac{\gamma_{Schw\Lambda}\nu}{T_{Schw\Lambda}-\nu}\right),
\eea
where $\gamma_{Schw\Lambda}=0.0397304$ and $\nu=-0.0138729$.  After normalizing the entropy such that the initial entropy is zero, we find that the entropy approaches a constant, as shown in Fig.(\ref{fig:entropyschwcosmo}).
We also wish to point out here that, though the effect is small, the entropy does in fact decrease for a short time before increasing.  One possible interpretation of this is that for this particular choice of parameters, the domain wall will not collapse spontaneously, but rather would require an energy input to get it started.  This conclusion is supported in the classical treatment by looking at Eq.($\ref{eq:Rdotnoapprox}$); some values of the cosmological constant (and therefore $f$), will yield a zero or negative number under the square root.  This would indicate that no collapse would occur.  The explanation of this effect is quite intuitive, as the collapse is at odds with the expansion of space, and if the cosmological constant is large enough, then collapse will be prevented.  We discovered that this feature is a function of the particular parameters that we chose.  The amount of entropy decrease is amplified when the domain wall starts closer to the cosmological horizon than the black hole horizon, and there is no longer a decrease in entropy when the cosmological constant is sufficiently small.  Given how small the cosmological constant is by current measurements, though, it is unlikely that the collapse of a de Sitter Schwarzschild domain wall would be forbidden in our universe (see for example \cite{Said:2012np}).  It should be noted, however, that the nature of entropy in the presence of a cosmological constant is currently not well understood, so our interpretation here is only speculative.

\subsection{(3+1) BTZ Domain Wall}
\begin{figure}
\centering
\includegraphics[width=3.2in]{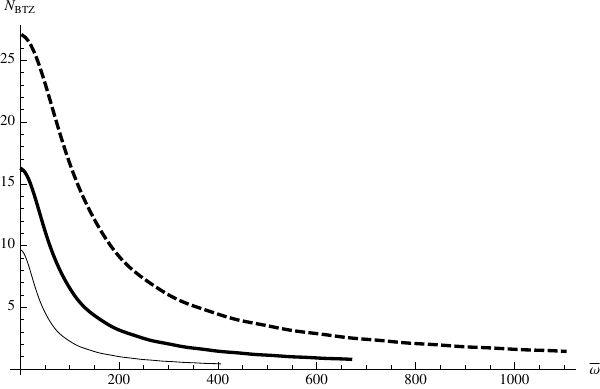}
\caption{Occupation number as a function of frequency for a BTZ domain wall.  From bottom to top, the three curves represent times slices for $t=13$ (solid), $t=14$ (bold), and $t=15$ (dashed), respectively.}
\label{fig:Nbtz}
\end{figure}

\begin{figure}
\centering
\includegraphics[width=3.2in]{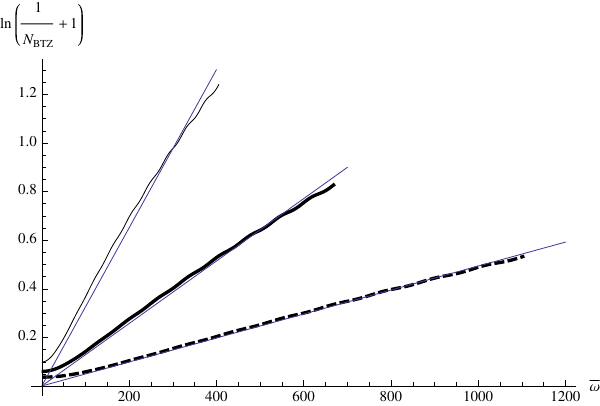}
\caption{The slope of this plot represents the thermodynamic quantity $\beta$ for a BTZ domain wall.  From top to bottom, the three curves represent times slices for $t=13$ (solid), $t=14$ (bold), and $t=15$ (dashed), respectively.}
\label{fig:lnNbtz}
\end{figure}

\begin{figure}
\centering
\includegraphics[width=3.2in]{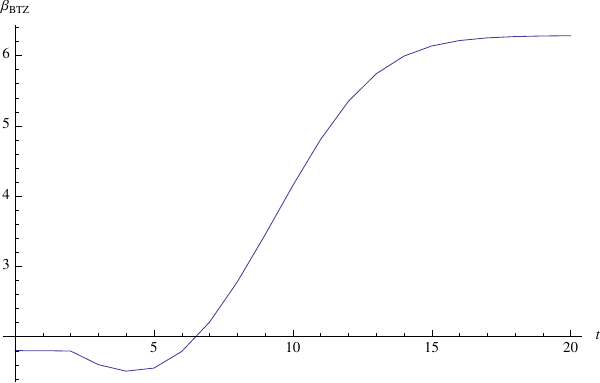}
\caption{$\beta$ as a function of time for a BTZ domain wall.}
\label{fig:betabtz}
\end{figure}

\begin{figure}
\centering
\includegraphics[width=3.2in]{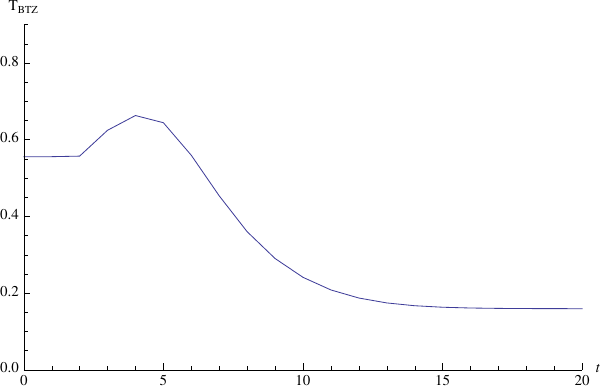}
\caption{Temperature as a function of time for a BTZ domain wall.}
\label{fig:tempbtz}
\end{figure}

Finally, we will determine the time evolution of the temperature and entropy of the (3+1) BTZ domain wall, for which
\be
f_{BTZ}=-\frac{4GM}{R}-\frac{\Lambda R^2}{3}\approx -\frac{4GM}{R}-\frac{\Lambda R_H^2}{3}
\ee
according to \cite{Lemos:1994xp}.
The plots of occupation number, $ln(1+1/N)$, $\beta$, and temperature are shown in Fig.(\ref{fig:Nbtz}), Fig.(\ref{fig:lnNbtz}), Fig.(\ref{fig:betabtz}), and Fig.(\ref{fig:tempbtz}), respectively.  Here we have used $\Lambda=-3$, where we have chosen the cosmological constant to be negative because there is no horizon otherwise.  

Once again, we will compare the late time temperature to the Hawking temperature.  According to \cite{Lemos:1994xp},
\be
T_H=\sqrt{-\frac{\Lambda}{3}}\frac{3}{4\pi}\left(\frac{M}{2}\right)^{1/3}.
\ee
As with the de Sitter Schwarzschild domain wall, the relationship between the horizon radius and the mass of the BTZ domain wall is not trivial.
\be
M=-\frac{\Lambda R^3}{12 G}
\ee

\begin{figure}
\centering
\includegraphics{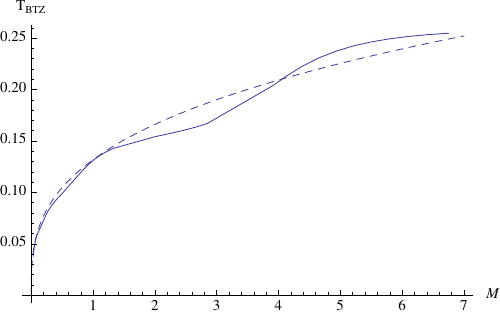}
\caption{Temperature as a function of mass for a BTZ domain wall.  The dashed line represents the Hawking model.}
\label{fig:tempdepbtz}
\end{figure}

The ratio of the late time temperature in the plot to the Hawking temperature for $R_H=1$ is $T_{BTZ}/T_H=1.33$.  The plot of the domain wall's late time temperature versus mass, as well as the best-fit curve, is shown in Fig.(\ref{fig:tempdepbtz}).  
The equation of the best-fit curve is
\be
T_{BTZ}=0.131724 M^{1/3}
\ee
After integrating, we find that the entropy as a function of temperature is 
\be
S_{BTZ}=\frac{3 T^2}{2 \gamma_{BTZ}},
\ee
where $\gamma_{BTZ}=0.131724$.

\begin{figure}
\centering
\includegraphics[width=3.2in]{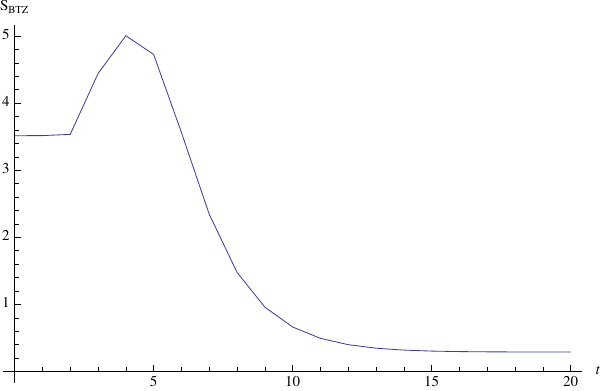}
\caption{Entropy as a function of time for a BTZ domain wall.}
\label{fig:entropybtz}
\end{figure}

The corresponding plot of entropy as a function of time is shown in Fig.(\ref{fig:entropybtz}).  
Here we have almost the opposite scenario from the de Sitter Schwarzschild domain wall.  In the beginning, the entropy increases as expected because when the domain wall is far enough from the event horizon it should collapse classically.  Right around the time at which the domain wall starts to slow down as it approaches the horizon, though, the entropy starts to decrease.  At late times, the domain wall reaches a static state and the entropy ceases to decrease.  Once again, one possible interpretation of the decreasing entropy is that the collapse would be prevented.  During the classical stage of the gravitational collapse, i.e. when the domain wall is farther from the horizon, the entropy of the system increases as expected.  As the domain wall approaches the horizon, it slows down significantly.  At some point, the gain in entropy from the collapse is overshadowed by the loss of entropy that results from the fact that the space is shrinking.  According to Ref\cite{Akbar:2011qw}, a BTZ black string does not evaporate once formed, and is thus stable.  However, our result of decreasing entropy seems to imply that a (3+1) BTZ domain wall would not completely collapse under these conditions, so the black string would never be formed in this way.  

\section{Conclusion}
\label{ch:conclusion}

We investigated the time evolution of the temperature and entropy of three types of domain walls.  This was done by coupling a scalar field to the background of the domain wall and evaluating the occupation number.  We first utilized a fully quantum approach, quantizing the position of the domain wall $R$.  From the occupation number we were able to determine $\beta$ and therefore the temperature.  We found that the late time temperature exhibited very good agreement with the Hawking temperature of a static black hole.  

We then repeated the process of determining the occupation number by coupling a scalar field to the background of the domain wall, but this time utilized a semi-classical approach.  For the Schwarzschild domain wall, the late time temperature was in extremely good agreement with the Hawking temperature, and the temperature scaled with the radius just as expected.  This allowed for an accurate determination of the entropy as a function of time.  It was found that the temperature decreased, then approached a constant.  The entropy, on the other hand, increased and approached a constant.

Results for the de Sitter Schwarzschild domain wall were very similar with a small exception.  The temperature first increased for a short time, then decreased and approached a constant.  It was found that this effect was suppressed for small values of the cosmological constant, so it appears that the difference in the time evolution was due to the cosmological constant.  The scaling of the temperature with the mass of the domain wall also matched the Hawking temperature very well.  The time evolution of the entropy exhibited an interesting feature in that it decreased for a short time.  This decrease was a function of the chosen parameters, and it implies that in order to collapse spontaneously, the domain wall might have to be given an initial input of energy.

The temperature and entropy of the (3+1) BTZ domain wall exhibited very interesting behavior, as they both increased initially, then decreased and approached a constant.  As was mentioned earlier, the de Sitter Schwarzschild domain wall also exhibited a decrease in entropy.  This implies that the decrease in entropy is related more to the cosmological constant than to the difference in topology.  The other interesting implication here, though, is that the collapse of a (3+1) BTZ domain wall will cease as it approaches the horizon radius.  Future work could further examine the expanding solutions.   

\section*{Acknowledgements}
The author would like to thank Peng Hao, Eric Greenwood, and Dejan Stojkovic for their contributions and input.

\end{document}